# *UMap* : Enabling Application-driven Optimizations for Page Management

Ivy B. Peng
peng8@llnl.gov
Lawrence Livermore National Laboratory
Livermore, USA

Marty McFadden
mcfadden8@llnl.gov
Lawrence Livermore National Laboratory
Livermore, USA

Eric Green
green77@llnl.gov
Lawrence Livermore National Laboratory
Livermore, USA

Keita Iwabuchi
iwabuchi1@llnl.gov
Lawrence Livermore National Laboratory
Livermore, USA

Kai Wu
kwu42@ucmerced.edu
University of California, Merced
Merced, USA

Dong Li
dli35@ucmerced.edu
University of California, Merced
Merced, USA

Roger Pearce
pearce7@llnl.gov
Lawrence Livermore National Laboratory
Livermore, USA

Maya Gokhale
gokhale2@llnl.gov
Lawrence Livermore National Laboratory
Livermore, USA

## ABSTRACT
Leadership supercomputers feature a diversity of storage, from node-local persistent memory and NVMe SSDs to network-interconnected flash memory and HDD. Memory mapping files on different tiers of storage provides a uniform interface in applications. However, system-wide services like mmap are optimized for generality and lack flexibility for enabling application-specific optimizations. In this work, we present *UMap* to enable user-space page management that can be easily adapted to access patterns in applications and storage characteristics. *UMap* uses the userfaultfd mechanism to handle page faults in multi-threaded applications efficiently. By providing a data object abstraction layer, *UMap* is extensible to support various backing stores. The design of *UMap* supports dynamic load balancing and I/O decoupling for scalable performance. *UMap* also uses application hints to improve the selection of caching, prefetching, and eviction policies. We evaluate *UMap* in five benchmarks and real applications on two systems. Our results show that leveraging application knowledge for page management could substantially improve performance. On average, *UMap* achieved 1.25 to 2.5 times improvement using the adapted configurations compared to the system service.

**Reference Format:**
Ivy B. Peng, Marty McFadden, Eric Green, Keita Iwabuchi, Kai Wu, Dong Li, Roger Pearce, and Maya Gokhale. 2019. *UMap* : Enabling Application-driven Optimizations for Page Management.

## 1 INTRODUCTION
Recently, leadership supercomputers provide enormous storage resources to cope with expanding data sets in applications. The storage resources come in a hybrid format for balanced cost and performance [9, 11, 13]. Fast and small storage, which is implemented using advanced technologies like persistent memory and NVMe SSDs, often co-locate with computing units inside compute node. Storage with massive capacity, on the other hand, uses cost-effective technologies like HDD and is interconnected to compute nodes through the network. In between, burst buffers use fast memory technologies and are accessible through the network. Memory mapping provides a uniform interface to access files on different types of storage as if to dynamically allocated memory. For instance, out-of-core data analytic workloads often need to process large datasets that exceed the memory capacity of a compute node [17]. Using memory mapping to access these datasets shift the burden of paging, prefetching, and caching data between storage and memory to the operating systems.

Currently, operating systems provide the *mmap* system call to map files or devices into memory. This system service performs well in loading dynamic libraries and could also support out-of-core execution. However, as a system-level service, it has to be tuned for performance reliability and consistency over a broad range of workloads. Therefore, it may reduce opportunities in optimizing performance based on application characteristics. Moreover, backing stores on different storage exhibit distinctive performance characteristics. Consequently, configurations tuned for one type of storage will need to be adjusted when mapping on another type of storage. In this work, we provide *UMap* to enable application-specific optimizations for page management in memory mapping various backing stores. *UMap* is highly configurable to adapt user-space paging to suit application needs. It facilitates application control on caching, prefetching, and eviction policies with minimal porting efforts from the programmer. As a user-level solution, *UMap* confines changes within an application without impacting other applications sharing the platform, which is unachievable in system-level approaches.

We prioritize four design choices for *UMap* based on surveying realistic use cases. First, we choose to implement *UMap* as a user-level library so that it can maintain compatibility with the fast-moving Linux kernel without the need to track and modify for frequent kernel updates. Also, we employ the recent userfaultfd [7] mechanism, other than the signal handling + callback function approach to reduce overhead and performance variance in multi-threaded applications. Third, we target an adaptive solution that sustains performance even at high concurrency for data-intensive applications, which often employ a large number of threads for hiding data access latency. Our design pays particular consideration on load imbalance among service threads to improve the utilization of shared resources even when data accesses to pages are skewed. *UMap* dynamically balances workloads among all service threads to eliminate bottleneck on serving hot pages. Finally, for flexible and portable tuning on different computing systems, *UMap* provides both API and environmental controls to enable configurable page sizes, eviction strategy, application-specific prefetching, and detailed diagnosis information to the programmer.

We evaluate the effectiveness of *UMap* in five use cases, including two data-intensive benchmarks, i.e., a synthetic sort benchmark and a breadth-first search (BFS) kernel, and three real applications, i.e., Lrzip [8], N-Store database [2], and an asteroid detection application that processes massive data sets from telescopes. We conduct out-of-core experiments on two systems with node-local SSD and network-interconnected HDD storage. Our results show that *UMap* can enable flexible user-space page management in data-intensive applications. On the AMD testbed with local NVMe SSD, applications achieved 1.25 to 2.5 times improvement compared to the standard system service. On the Intel testbed with network-interconnected HDD, *UMap* brings the performance of the asteroid detection application close to that uses local SSD for 500 GB data sets. In summary, our main contributions are as follows:

- We propose an open-source library[1], called *UMap* that leverages lightweight userfaultfd mechanism to enable application-driven page management.
- We describe the design of *UMap* for achieving scalable performance in multi-threaded data-intensive applications.
- We demonstrate five use cases of *UMap* and show that enabling configurable page size is essential for performance tuning in data-intensive applications.
- *UMap* improves the performance of tested applications by 1.25 to 2.5 times compared to the standard mmap system service.

## 2 BACKGROUND AND MOTIVATION

In this section, we introduce memory mapping, prospective benefits from user-space page management, and the enabling mechanism *userfaultfd*.

### 2.1 Memory Mapping

Memory mapping links images and files in persistent storage to the virtual address space of a process. The operating system employs demand paging to bring only accessed virtual pages into physical memory because virtual memory can be much larger than physical memory. An access to memory-mapped regions triggers a page fault if no page table entry (PTE) is present for the accessed page. When such a page fault is raised, the operating system resolves it by copying in the physical data page from storage to the in-memory page cache.

Common strategies for optimizing memory mapping in the operating systems include page cache, read-ahead, and madvise hints. The page cache is used to keep frequently used pages in memory while less important pages may need to be evicted from memory to make room for newly requested pages. Least Recently Used (LRU) policy is commonly used for selecting pages to be evicted. The operating system may proactively flush dirty pages, i.e., modified pages in the page cache, into storage when the ratio of dirty page exceeds a threshold value [19]. Read-ahead preloads pages into physical memory to avoid the overhead associated with page fault handling, TLB misses and user-to-kernel mode transition. Finally, the madvise interface takes hints to allow the operating system to make informed decisions for managing pages.

---

[1]UMAP v2.0.0 https://github.com/LLNL/umap

## 2.2 User-space Page Management

User-space page management uses application threads to resolve page faults and manage virtual memory in the background as defined by the application. The userfaultfd is a lightweight mechanism to enable user-space paging compared to the traditional *SIGSEGV* signal and callback function [7]. Applications register address ranges to be manged in user-space, and specify the type of events, e.g., page faults and events in un-cooperative mode, to be tracked. Page faults in the address ranges are delivered asynchronously so that the faulting process is blocked instead of idling, allowing other processes to be scheduled to proceed.

The fault-handling thread in the application can atomically resolve page faults with the *UFFDIO_COPY* ioctl, which ensures the faulting process is (optionally) waken up only after the requested page has been fully copied into physical memory [7]. The fault-handling threads may utilize application-specific knowledge to optimize this procedure, providing the flexibility that is unachievable in kernel mode. For instance, the application could select arbitrary page sizes, read-ahead window size, or provides specific pages for prefetching or evicting. All these optimizations remain inside one application and will not impact other applications sharing the same system. User-space paging is not only limited to backing store on file systems. In contrast to kernel mode, the fault-handling thread has the liberty to fetch data from a variety of backing stores, such a memory server, databases, and even another process.

## 3 DESIGN

In this section, we describe the design of *UMap* . We first provide an overview of the architecture and then focus on four optimizations for achieving high performance in user-space.

### 3.1 Overview

*UMap* provides an interface for applications to register multiple virtual address ranges, called *UMap regions* that bypass the kernel service and instead, be managed in user-space. Figure 1 presents the *UMap* architecture. Dark blue regions in the virtual address space are *UMap* regions. Each region has a backing store, where the data is physically located. *UMap* provides an abstraction layer in the store object (yellow circles) for accessing different types of storage. When an application accesses a *UMap* region, if the accessed page is not present in the physical memory, page faults are triggered. These page faults queue up in a FIFO buffer and multiple *UMap* fillers cooperatively resolve these faults. If the requested pages are not fetched in yet, *UMap* fillers will invoke the access functions defined in the store object to read data from the underlying storage. If the buffer is fully occupied, some pages need to be evicted following a user-defined strategy. In the background, a group of *UMap evictors* keep monitoring the ratio of dirty pages in the buffer. Once the ratio of dirty pages reaches a (configurable) high watermark, *UMap* evictors will coordinately write data to the storage.

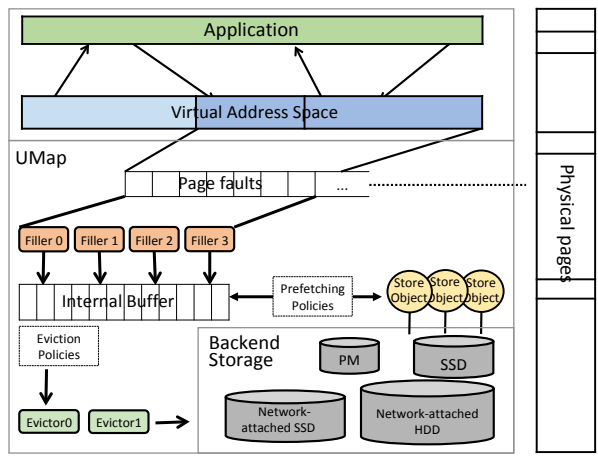

**Figure 1: The *UMap* architecture.**

### 3.2 I/O Decoupling

Our design decouples the I/O operation from the fault-handing threads to achieve high concurrency in long latency tasks. I/O operations that move data between storage and memory have a much longer latency than memory accesses. For instance, latency to the state-of-art persistent memory (PM) is about 100 - 500 ns [12], latency to NVMe-based SSD is in the range of $\approx 20\ \mu s$ [3] while accesses to HDD would require several milliseconds. In contrast, memory accesses typically takes 20-100 ns. To improve the I/O performance, *UMap* employ a configurable number of threads for moving data between storage and memory to exploit the bandwidth supported by the hardware.

The dedicated two groups of I/O threads is referred to as *fillers* and *evictors*, as illustrated in the orange and blue boxes in Figure 1. Fillers split the workload of copying pages to memory while evictors concurrently write data to storage. A separate group of manager threads, typically with low concurrency, keeps polling for notification of tracked events from the operating system. By decoupling the tasks into three groups of workers, *UMap* has the flexibility to adapt the concurrency in each group to reflect their different workload. In contrast, a coupled design results in a long blocking operation that has limited flexibility to optimize.

### 3.3 Dynamic Load Balancing

*UMap* employs a dynamic load balancing strategy to improve resource utilization. We find that memory-mapped regions could have hot and cold segments. Hot segments require a higher level of concurrency for frequent data movement and more physical memory for buffering data than cold segments. For instance, social networks are considered as a type of scale-free network whose degree distribution follows a power law. Memory segment that stores high-degree vertices would naturally result in more accesses than the regions that store low-degree vertices. We design *UMap* to avoid load imbalance even in such skewed data access patterns by dynamically distributing workloads from all memory regions among *UMap* fillers.

*UMap* employs a dynamic scheduling strategy similar to "work stealing" approach in task-based programming models [15]. *UMap* uses a single *UMap buffer* object to manage the metadata of in-memory pages for all regions. When *UMap* receives the notification of a fault event from the operating system, it appends the workload for resolving this fault into a dynamically growing queue. A group of workers split the pending workload to load pages from the backing store collectively. Consequently, when hot memory segments generate more workloads than others, they will be assigned with more working threads. Orthogonal to the data fetching task is the data flushing task that writes dirty pages back to the persistent stores. When the number of dirty pages reaches a high watermark, the workload is appended to a separate queue and then split by a different group of workers. Figure 1 illustrates the shared (internal) buffer and the work distribution among workers. The dynamic load balancing design prepare *UMap* to cope with applications with diverse access patterns.

### 3.4 Extensible Back Store

*UMap* provides a data object abstraction layer to support different types of backing stores. Currently, applications running on leadership supercomputers have multiple choices of storage, including local SSD, network-interconnected SSD, and HDD. In the future, architectures that provide disaggregated memory and storage resources are likely to emerge. Based on this observation, our design ensures that *UMap* is extensible for current and future architectures.

*UMap* facilitates applications to associate their own backing store for each memory region. The application has specific control over which storage layer to access to resolve a page fault. In this way, an application is presented with a uniform interface as the virtual memory address space while *UMap* in the backend handles data movement to/from various types of storage.

### 3.5 User-controlled Page Flushing

We design *UMap* to enable user-space control on page flushing to a persistent store. There are two motivations. First, the system service may write dirty pages to storage whenever the operating system deems appropriate. Unpredictable behavior may occur if a memory range requires strong consistency such as atomicity among multiple pages. Second, frequent page flushing is known to cause increased performance variation and degradation. For instance, RHEL trigger page flushing when more than 10% pages are dirty [19]. With user control, the application could avoid aggressive page flushing by setting a high threshold or even postponed page flushing to a later stage. *UMap* monitors the ratio of dirty pages to compare with a user-defined high watermark to trigger page flushing as well as a low watermark that suspends page flushing.

### 3.6 Application-Specific Optimization

*UMap* maintains a set of parameters for programmers with application knowledge to configure page management. One of the most performance-critical parameters is the internal page size of a memory region, denoted as *UMap* page. *UMap* supports an arbitrary page size for each memory region while the system service only supports fixed page sizes. *UMap* page defines the finest granularity in data movement between memory and backing store. For the same memory region, choosing a large *UMap* page could reduce the overhead of metadata, but may also move more than accessed data into memory. By tuning the page size, an application could identify an optimal configuration that balances the overhead and data usage. Also, an application can control the page buffer size, which can alleviate OOM situations in unconstrained mmap.

*UMap* also supports a flexible prefetching policy that can fetch pages even in irregular patterns. The operating systems usually recognize page accesses as either sequential or random, to increase or decrease the readahead window size, respectively. Real-world applications, however, exhibit complex access patterns, and the general prefetching mechanism becomes insufficient. In contrast, *UMap* could prefetch a set of arbitrary pages into memory, as informed by the application. Moreover, an application can control the start of prefetching to avoid premature data migration that interferences with pages in use. This flexibility, together with knowledge from application algorithm or offline profiling, eases application performance tuning.

## 4 IMPLEMENTATION

*UMap* is implemented in C++ and uses the userfaultfd system call [1]. *UMap* enables application controls on page management through both API and environmental variables. The fault-handling thread resolves the page fault by calling the application-supplied function (if provided), or performing direct I/O to the backing store by invoking the defined access functions. *UMap* uses the UFFDIO_COPY ioctl [7] to ensure atomic copy to the allocated memory page before waking up the blocked process.

### 4.1 API

*UMap* provides similar interfaces as mmap to ease porting existing applications. An application can register/unregister multiple memory regions to be managed by *UMap* through the umap and uunmap interface. One additional flexibility provided by *UMap* is the multi-file backed region. Given a set of files, each with individual offsets and size, *UMap* maps them into a contiguous memory region. While applications can rely on *UMap* runtime for managing pages, *UMap* also provides a plugin architecture that allows application to register callback functions. A set of configuration interfaces with naming convention umapcfg_set_xx, allow the application to control paging explicitly: (1) the maximum size of physical memory used for buffering pages; (2) the level of concurrency for processing I/O operations in each group of workers; (3) the threshold value for starting or suspending writing dirty pages to back stores. Listing 1 illustrates a simple application that uses paging and prefetching services in *UMap* .

**Listing 1: UMap API**

```
int fd = open(fname, O_RDWR);
void* base_addr = umap(NULL, totalbytes,
    PROT_READ|PROT_WRITE, UMAP_PRIVATE, fd, 0);

//Select two non-contiguous pages to prefetch
std::vector<umap_prefetch_item> pfi;
umap_prefetch_item p0 = { .page_base_addr = &base[5 *
    psize] };
pfi.push_back(p0);
umap_prefetch_item p1 = { .page_base_addr = &base[15
    * psize] };
pfi.push_back(p1);
umap_prefetch(num_prefetch_pages, &pfi[0]);

computation();

//release resources
uunmap(base_addr, totalbytes);
```

### 4.2 Environmental Controls

*UMap* uses a set of environment variables to control: the number of fillers and evictors; the buffer size; the buffer draining policy; and the read-ahead window size. We highlight the key environment variables that *UMap* tracks to dictate its runtime behavior:

• UMAP_PAGESIZE sets the internal page size for memory regions

• UMAP_PAGE_FILLERS sets the number of workers to perform read operations from the backing store. Default: the number of hardware threads.

• UMAP_PAGE_EVICTORS sets the number of evictors that will perform evictions of pages. Eviction includes writing to the backing store if the page is dirty and informing the operating system that the page is no longer needed. Default: the number of hardware threads.

• UMAP_EVICT_HIGH_WATER_THRESHOLD sets the threshold in *UMap* buffer to trigger the evicting procedure. Default: 90%

• UMAP_EVICT_LOW_WATER_THRESHOLD sets the threshold in *UMap* buffer to suspend evicting procedure. Default: 70%

• UMAP_BUFSIZE sets the size of physical memory to be used for buffering *UMap* pages. Default: (80% of available memory)

• UMAP_READ_AHEAD sets the number of pages to read-ahead when resolving a demand paging. Default: 0

• UMAP_MAX_FAULT_EVENTS: sets the maximum number of page fault events that will be read from the kernel in a single call. Default: the number of hardware threads.

### 4.3 Limitations

The current implementation uses the write protection support from the kernel to track dirty pages in the physical memory. For pages in write-protected memory ranges, a writes will trigger a fault that sends a UFFD message to handling threads. Currently, the write protection support in userfaultfd is only available in the experimental Linux kernel [2].

## 5 EXPERIMENTAL SETUP

In this section, we describe the experimental setup for the evaluation. We summarize the configuration parameters of two testbeds in Table 1 and 2. The AMD testbed includes three identical machines (Altus, Bertha, Pmemio) that feature two AMD EPYC 7401 (24 cores /48 hardware threads) processors. The testbed has a total of 256 GB DDR4 DRAM and 16 memory channels that operate at 2400 MT/s. Each machine has a total of 4.65 TiB disk capacity, including 1.8 GB SATA Micron 5200 Series SATA SSD. The platform runs Fedora 29 with Linux kernel 5.1.0-rc4-uffd-wp-207866-gcc66ef4-dirty (experimental version) . We compiled all applications using GCC 8.3.1 compiler with support for OpenMP. We use the local SSD on the AMD testbed to evaluate the impact of *UMap* page sizes in all applications. The second testbed, the Intel testbed is on a cluster called *flash*. Its storage includes a remote HDD through Lustre parallel distributed file system. It also features 1.5 TB local SSD. We test the asteroid detection application on this testbed to compare the performance of the backing store on Lustre with the local SSD. The platform runs the Red Hat Enterprise Linux 7.6 kernel. We compiled all applications using GCC 8.1.0 compiler.

---
[2]Linux Patch https://git.kernel.org/pub/scm/linux/kernel/git/andrea/aa.git.

Table 1: The AMD Testbed Specifications

| Platform | Penguin® Altus® XE2112 (Base Board: MZ91-FS0-ZB) |
|---|---|
| Processor | AMD EPYC 7401 |
| CPU | 24 cores (48 hardware threads) × 2 sockets |
| Speed | 1.2 GHz |
| Caches | 64KB 8-way L1d and 32KB 4-way L1i, 512KB 8-way private L2, 8MB 8-way shared L3 per three cores |
| Memory | 16 GB DDR4 RDIMM × 8 channels (2400 MT/s) × 2 sockets |
| Storage | ≈ 3 TB NVMe (type: HGST SN200) |

Table 2: The Intel Testbed Specifications

| Platform | S2600WTTR (Base Board: S2600WTTR) |
|---|---|
| Processor | Intel Xeon E5-2670 v3 (Haswell) |
| CPU | 12 cores (24 hardware threads) × 2 sockets |
| Speed | 2.3 GHz (Turbo 3.1 GHz) |
| Caches | 32KB 8-way L1d and 32KB 8-way L1i, 256KB 8-way private L2, 30MB 20-way shared L3 |
| Memory | 2 16 GB DDR4 RDIMM × 4 channels (1866 MT/s) × 2 sockets |
| Storage | ≈ 1.5 TB NVMe SSD(type: HGST SN200) |

## 6 EVALUATION

In this section, we evaluate the performance of *UMap* in data-intensive benchmarks and applications. In particular, we study the performance benefit of enabling flexible page sizes at application level.

### 6.1 Out-of-core Sort

Our first evaluation uses an in-house sorting benchmark, called umapsort. Umapsort is a multi-threaded program that performs quicksort on values stored in a file. Thus, umapsort is a read-write workload. For the evaluation, we use a single 500GiB data set of a sequence of ascending 64-bit words. We configured the benchmark to memory map data sets either using the mmap system call or *UMap* API. Then, the program sorts the values in the memory region into descending order. The application was configured to run with 96 OpenMP threads on the AMD testbed with 256GiB of physical memory. The data set is stored on the local NVMe-SSD device configured with its default boot-time values. We report the experimental results in Figure 2.

We used different numbers of fillers and evictors to identify the optimal concurrency for this benchmark. In most tested cases, using 48 fillers and 24 evictors brings the best performance. We then fixed the number of fillers and evictors to test the impact of different page sizes. For the mmap tests, we use its default setting and the standard 4KiB page size.

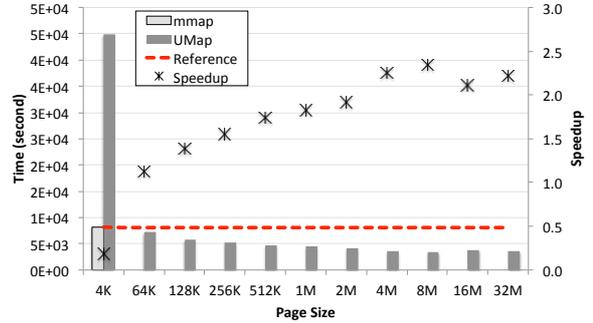

Figure 2: The performance of *UMap* for sorting 500 GiB data on NVMe-SSD on the AMD testbed, as normalized to that of mmap. *UMap* starts outperforming mmap when the page size is larger than 64KB. At the page size of 8 MB, *UMap* achievs 2.5 times improvement compared to mmap.

For *UMap* tests, we change the page size to identify the optimal configuration. At the smallest page size, *UMap* shows much higher overhead than mmap. We find that increasing page sizes in *UMap* steadily improves the performance. At 64KiB page, *UMap* starts outperforming mmap. By adjusting *UMap* page size to 8MiB, the *UMap* version achieves 2.5 times speedup compared to the mmap version. One reason for the improved performance at larger page sizes is that the reduction in page faults, which reduces the time spent in servicing page faults and also aggregate smaller data transfers into bulky transfers to exploit bandwidth. As the change is localized to the application process, there is no need to modify any OS page size or file system prefetch settings.

### 6.2 Graph Application

We implemented a conventional level-synchronous BFS algorithm. Our BFS program takes a graph with compressed sparse row (CSR) data format and stores only the CSR graph in the storage device. We used a separated program to generate a CSR graph to make a read-only benchmark and dropped page cache before running the benchmark to achieve consistent results. As for dataset, we used an R-MAT graph generator with the edge falling probabilities used in the Graph500.

Figure 3 shows Umap's BFS performance normalizing to mmap's best performance case where readahed is off. We varied Umap page size from 4 KB to 4 MB and used the default values for its other environmental variables. Umap showed its best performance and overperformed mmap by 1.8X with 512 KB page size whereas mmap slowed down as increased the page size. We clearly confirmed the benefit of Umap's variable page size feature in terms of not only providing user level control but also better performance.

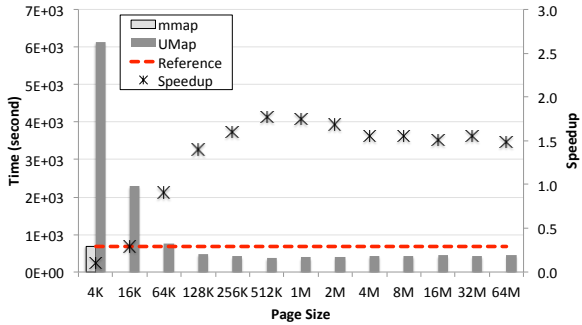

Figure 3: The relative performance of *UMap* as compared to that of mmap in BFS on an R-MAT scale 31 CSR graph (529 GB) data on NVMe on the AMD testbed.

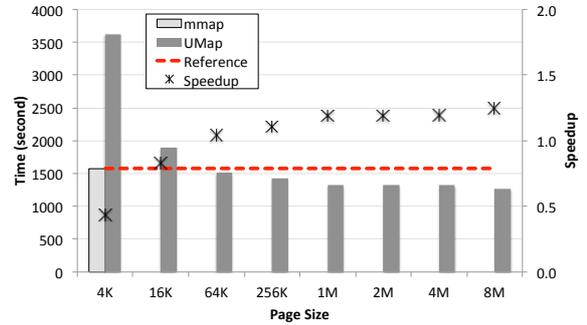

Figure 4: The relative performance of *UMap* as compared to that of mmap version for LRZIP 64 GB random data on NVMe on the AMD testbed.

### 6.3 File Compression

Long Range ZIP (lrzip) is a program that implements a full-file compression algorithm [8]. Compression algorithms detect redundancies in input files to reduce size. Lrzip uses a modified RZIP algorithm to achieve an effectively unlimited compression window size. The original mmap version of lrzip uses a large buffer, e.g., one-third of system memory, to mmap a window that 'slides' through the input file. When matches are found, lrzip may use a secondary 64k mmap region to page in any matching regions outside the main window. The *UMap* version removes these sliding buffers and replaces them with a single *UMap* region spanning the entire input file. *UMap* runtime automatically manages the amount of file data paged in memory during execution.

Our experiments run lrzip in pre-processing mode to compare the performance of mmap with *UMap* in RZIP algorithm. We constrain the available memory to the program to ensure out-of-core execution, i.e., 16 GB memory and a 64 GB input data. The *UMap* version sets the environmental variable to limit *UMap* buffer for caching pages in memory. The mmap version requires a command-line option to override the system memory on the testbed. In Figure 4 , lrzip shows low sensitivity to the change in page size. This insensitivity is likely due to the mostly sequential access pattern in lrzip, which only has occasional data reuse of earlier portions of the input file, i.e., when duplicated hash values are found. Once the page size exceeds 1MB, the *UMap* version stabilizes performance at about 1.25 times that of the mmap system call.

### 6.4 Asteroid Detection Application

In this case study, we use *UMap* for an on-going study that searches for transient objects, such as asteroids, in intermittent time-series telescope data. We uses UMap to create a 3D cube of virtual address space, where each page is directly mapped to pixel data in a series of image files. UMap has the extensibility to integrate an application-specific FITS handler for resolving page fault to a particular file, which would require extensive porting efforts to achieve in mmap.

The application creates millions or even billions of vectors and then virtually traces them through the image cube to calculate the median pixel value along each vector. The starting point of each vector has a uniform random distribution in the data and their slope follows a given linear function. The backing store contains thousands of FITS format image files. Page faults are resolved to the FITS files containing the requested data, where the pixel data is subsequently read and decoded before copied into the faulting page. Note that a page fault may require access data in multiple files.

The evaluation uses a synthetic data set derived from 537 random images taken from an astronomical survey performed on 12/232018 by the Dark Energy camera in Chile. These files were resized via bicubic resampling to four times their original dimension in each axis in order to emulate the characteristics of real-world datasets. Each file is approximately 977MB with dimensions of 16,000 by 16,000 pixels after this operation. The entire dataset is approximately 512GB. For the Lustre tests, transparent Lustre compression and de-duplication reduces this size to 223GB.

The experiments process a single pass of 32 million vectors with a *UMap* buffer size of 64GB. We demonstrate two types of backing stores in this application. The first uses the local SSD on the AMD testbed. The second uses a backing store mapped to remote disks through a Lustre parallel file system on the Intel testbed. Figure 5 and 6 present the results. Our results show that the application has low sensitivity to page sizes because data reuse among the vectors. A slight performance degradation at large page sizes because larger pages bring more unused data. The execution time initially decreases to the optimal minimum at 1MiB page and then, slightly increases as larger amounts of unused data begins to contend for buffer space.

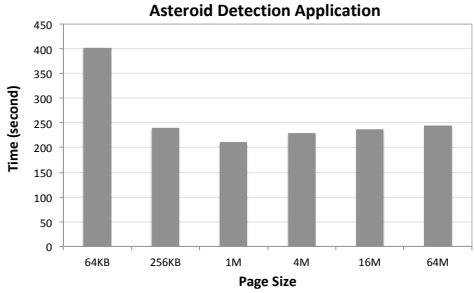

Figure 5: Execution time of the asteroid application on local SSD at various *UMap* page sizes at 256GB input.

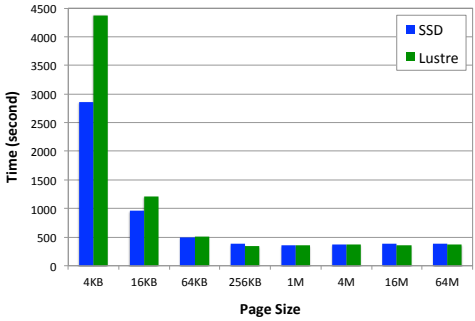

Figure 6: Compare performance of the asteroid application on local SSD and Lustre using 512GB input.

### 6.5 Database Workload

This use case demonstrates that *UMap* can be easily plugged into existing database applications to improve user-space control over memory mapping. We ported N-Store [2], an efficient NVM database, to use *UMap* API by changing approximately ten lines of code. N-Store uses persistent memory like SSD as the memory pool for data. Our experiments use a 384 GB persistent memory pool on the local NVMe-SSD on the AMD testbed. N-Store supports multiple executors to execute transactions to the database concurrently. In our evaluation, we sweep 4-32 executors to understand the scalability of *UMap* on variable concurrency. Our workload uses the popular YCSB [4] benchmark with eight million transactions and five million keys. The measurement is repeated ten times, and we report throughput from N-Store as the metric for performance.

We tested different numbers of fillers and evictors to select the concurrency to be 48 fillers and 24 evictors for this benchmark. Then, with a fixed number of fillers and evictors, we test the impact of different page sizes. Figure 7 reports the throughput of *UMap* version at different page sizes and the original mmap version at the default 4KiB page. We find that increasing page sizes in *UMap* does show a trend of increased performance as other applications. The highest throughput is achieved at 32KiB page size, which is about 34% improvement of the mmap version. This page size is smaller than the optimal page sizes in other applications because the access pattern in the benchmark has low locality and mostly random.

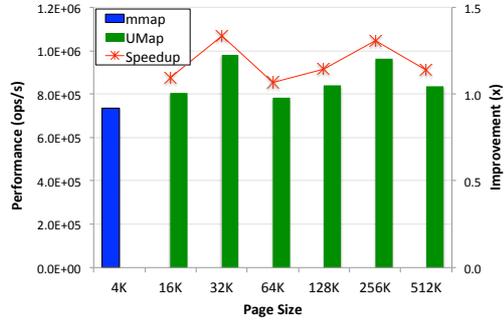

Figure 7: Compare database throughput using mmap and *UMap*. *UMap* achieves up to 34% improvement at 32KB page.

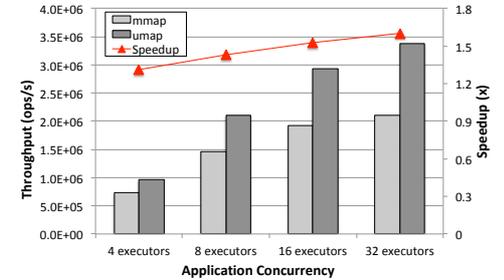

Figure 8: A scaling test in N-Store using increased number of executors in the database shows that UMap sustains performance scaling at increased application concurrency.

Figure 8 report the throughput of the database at an increased application concurrency, i.e., the number of executors increases. The scaling test results demonstrate the advantage of *UMap* in addressing application requirements that change dynamically. When the number of executors increases from four to 32, the gap between the *UMap* version and the mmap version increases (in the gray bars). In particular, the speedup by *UMap* increases from 1.3x to 1.6x steadily (the red line). This result highlights the importance of a scalable design in *UMap* for handling various application workloads.

## 7 DISCUSSION

There are several future directions for *UMap* to support emerging architectures.

**Multi-tiered Storage** has tiered access latency and bandwidth. Currently, *UMap* is extensible for new layers by defining new data objects. In the future work, we will automate data migration between data objects and adapt to application characteristics to improve storage utilization.

**Disaggregated Memory** architecture has large-capacity memory servers connected to compute node through high-performance network to provide memory on demand. *UMap* can be used to port applications on such architecture by providing a backing store that defines access functions likely using RDMA for moving to/from memory server.

**Byte-addressable NVM** requires strong consistency for system software like file systems and DAX-aware mmap lacks such support [20]. The *UMap* buffer could provide applications with explicit control on when to persist changes cached in volatile memory.

## 8 RELATED WORKS

Previous works have identified limitations in system services for data-intensive applications that perform out-of-core execution for large data sets [5, 18]. [16] analyzes the overhead in the path through Linux virtual memory subsystem for handling memory-mapped I/O. They conclude that kernel-based paging will prevent applications to exploit fast storage. Our approach aims to provide flexibility to adapt memory mapping to application characteristics and back store features.

DI-MMAP [17] provides a loadable kernel module that combines with a runtime to optimize page eviction and TLB performance. This approach requires updates to remain compatible with the fast-moving kernel. CO-PAGER [10] also provides a user-space paging service by combining a kernel module with a user-space component. CO-PAGER bypasses complex I/O subsystem in the kernel to reduce the overhead of accessing NVM. Our approach stays in user-space completely, and require no modification in the kernel or updates due to kernel updates. Moreover, our design can support a variety of back stores. For instance, remote memory paging that fetches data from a memory server or compute node [6, 14] could be easily integrated into *UMap* by providing a new store object.

## 9 CONCLUSIONS

In this work, we provide a user-space page management library, called *UMap* , to flexibly adapt memory mapping to application characteristics and storage features. *UMap* employs the lightweight userfaultfd mechanism to enable applications to control critical parameters that impact the performance of memory mapping large data sets while confining the customizations within the application without impacting other applications on the same system. We evaluate *UMap* in five applications using large data sets on both local SSD and remote HDD. By adapting the page size in each application, *UMap* achieved 1.25 to 2.5 times improvement compared to the system service mmap. In summary, ă*UMap* can be easily plugged into data-intensive applications to enable application-specific optimization.

## ACKNOWLEDGMENT

This work was performed under the auspices of the U.S. Department of Energy by Lawrence Livermore National Laboratory under Contract DE-AC52-07NA27344 (LLNL-PROC-788145). This research was also supported by the Exascale Computing Project (17-SC-20-SC), a collaborative effort of the U.S. Department of Energy Office of Science and the National Nuclear Security Administration. This document was prepared as an account of work sponsored by an agency of the United States government. Neither the United States government nor Lawrence Livermore National Security, LLC, nor any of their employees makes any warranty, expressed or implied, or assumes any legal liability or responsibility for the accuracy, completeness, or usefulness of any information, apparatus, product, or process disclosed, or represents that its use would not infringe privately owned rights. Reference herein to any specific commercial product, process, or service by trade name, trademark, manufacturer, or otherwise does not necessarily constitute or imply its endorsement, recommendation, or favoring by the United States government or Lawrence Livermore National Security, LLC. The views and opinions of authors expressed herein do not necessarily state or reflect those of the United States government or Lawrence Livermore National Security, LLC, and shall not be used for advertising or product endorsement purposes.